%% file: paper.tex
\documentclass[11pt]{article}
\usepackage[colorlinks=true,linkcolor=blue,urlcolor=black,citecolor=blue,breaklinks]{hyperref}
\usepackage{color}
\usepackage{enumerate}
\usepackage{graphicx}
\usepackage{varwidth}
\usepackage{mathrsfs}
\usepackage{mathtools}
\usepackage[font=small,labelfont=bf]{caption}
\usepackage{subcaption}
\usepackage{overpic}
\usepackage{multirow}
\usepackage{natbib}
\usepackage{tikz}
\bibpunct{(}{)}{;}{a}{,}{,}

\usepackage{fullpage}
\usepackage{authblk}
\usepackage{amsmath,amsthm,amssymb,colonequals,etoolbox}
\usepackage{thmtools}
\usepackage{url}
\usepackage[capitalize,noabbrev]{cleveref}
\crefformat{equation}{(#2#1#3)}
\crefrangeformat{equation}{(#3#1#4) to (#5#2#6)}
\crefmultiformat{equation}{(#2#1#3)}{ and (#2#1#3)}{, (#2#1#3)}{ and (#2#1#3)}

\input{commands}
\title{Model-free learning of probability flows:\\Elucidating the nonequilibrium dynamics of flocking}

\author[1]{Nicholas M.~Boffi}
\affil[1]{Carnegie Mellon University}
\author[2]{Eric Vanden-Eijnden}
\affil[2]{Courant Institute of Mathematical Sciences}

\begin{document}
\maketitle

\begin{abstract}
\input{abstract}
\end{abstract}

\input{intro}
\input{problem}
\input{results}
\input{conc}

\bibliographystyle{unsrtnat}
\bibliography{paper}

\input{end_matter}

\clearpage
\appendix
\input{app}
\end{document}

%% file: commands.tex
\newcommand{\calL}{\mathcal{L}}
\newcommand{\calP}{\mathcal{P}}

\newcommand{\R}{\ensuremath{\mathbb{R}}}

\newcommand{\norm}[1]{\lVert #1 \rVert}

\newcommand{\E}{\mathbb{E}}

\newcommand{\normal}{\mathsf{N}}

\newcommand{\sign}{\mathsf{sign}}

\newcommand{\kl}[2]{\mathsf{KL}(#1\:\Vert\:#2)}

\newcommand{\Sdottot}{\dot{S}_{\text{tot}}}
\newcommand{\sdottot}{\dot{s}_{\text{tot}}}

\newcommand{\sdotsys}{\dot{s}_{\text{sys}}}

\newcommand{\rev}{\mathsf{R}}
\newcommand{\tot}{\text{tot}}
\newcommand{\stoch}{\text{stoch}}
\newcommand{\dt}{\mathrm{d}t}

%% file: abstract.tex
Active systems comprise a class of nonequilibrium dynamics in which individual components autonomously dissipate energy.
Efforts towards understanding the role played by activity have centered on computation of the entropy production rate (EPR), which quantifies the breakdown of time reversal symmetry.
A fundamental difficulty in this program is that high dimensionality of the phase space renders traditional computational techniques infeasible for estimating the EPR.
Here, we overcome this challenge with a novel deep learning approach that estimates probability currents directly from stochastic system trajectories.
We derive a new physical connection between the probability current and two local definitions of the EPR for inertial systems, which we apply to characterize the departure from equilibrium in a canonical model of flocking.
Our results highlight that entropy is produced and consumed on the spatial interface of a flock as the interplay between alignment and fluctuation dynamically creates and annihilates order.
By enabling the direct visualization of when and where a given system is out of equilibrium, we anticipate that our methodology will advance the understanding of a broad class of complex nonequilibrium dynamics.

%% file: intro.tex
\section{Introduction.}
\label{sec:intro}

Active matter systems are driven out of equilibrium at the microsopic level as their individual agents consume free energy and perform work on their environment~\citep{ramaswamy_mechanics_2010, marchetti_hydrodynamics_2013, bowick_symmetry_2022}.
The nonequilibrium character of their dynamics manifests itself in the breakdown of time-reversal symmetry (TRS)~\citep{lebowitz_gallavotticohen-type_1999}, whose significance can be quantified by the entropy production rate (EPR)~\citep{seifert_entropy_2005, seifert_stochastic_2012}.
As a high-dimensional quantity that depends on the unknown nonequilibrium stationary density, estimation of the EPR is of considerable research interest~\citep{fodor_how_2016}.
Several recent thrusts in this direction include the development of analytical tools rooted in active field theories~\citep{nardini_entropy_2017}, information-theoretic approaches~\citep{ro_model-free_2022, martiniani_quantifying_2019}, and computational methods based on machine learning~\citep{boffi_deep_2024, boffi_probability_2023}.
Thus-far, machine learning-based efforts have required knowledge of the system's dynamics, which is unavailable for many physical~\citep{ballerini_empirical_2008} and experimental systems of interest~\citep{palacci_living_2013}.

In this Letter, we introduce a novel machine learning algorithm that estimates the EPR from system trajectories alone, obviating the need for knowledge of the dynamics and paving the way for applications to experimental systems.
To illustrate the utility of our approach, we apply the method to understand the nonequilibrium dynamics of flocking~\citep{toner_flocks_1998, toner_hydrodynamics_2005, toner_long-range_1995} in a Vicsek-like model~\citep{vicsek_novel_1995, chate_modeling_2008}, a canonical active matter system.
In particular, we compute spatial decompositions of both the total EPR and the system EPR in the sense of Seifert~\citep{seifert_entropy_2005,boffi_deep_2024}.
To our knowledge, this is the first study that provides a spatial decomposition of the breakdown of TRS in flocking models.

Our findings reveal that TRS violations concentrate on the interface between the flock and the disordered gas, and that the EPR of the system gives insight into each particle's instantaneous contribution to the generation or annihilation of order.
In addition, we show that TRS breaking events exhibit intermittent dynamics~\citep{hirsch_theory_1982} reflective of the formation and breakdown of flocks, which we quantify by identifying $1/f$ noise in the EPR~\citep{manneville_intermittency_1980}.

%% file: problem.tex
\section{Problem setting}
\paragraph{Microscopic model.}
\begin{figure}[!t]
	\centering
	\begin{overpic}[width=0.4\textwidth]{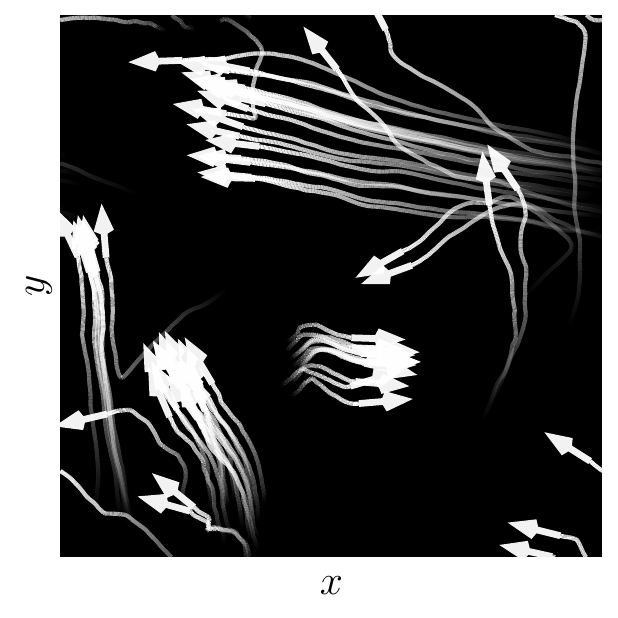}
	\end{overpic}
	\caption{
		\textbf{The nonequilibrium dynamics of flocking.}
		System dynamics for~\eqref{eqn:vicsek}, with trail indicating past motion.
		The particles spontaneously self-organize into polar flocks that irregularly split and combine in a manner that breaks time reversal symmetry.
		In this work, we study how these TRS breaking events are distributed spatially.}
	\label{fig:overview}
\end{figure}
We study a system of $N$ interacting particles motivated by the Vicsek model of flocking~\citep{vicsek_novel_1995,chate_modeling_2008}. %
Denoting by $x_t= (x^1_t,x^2_t, \ldots, x^N_t)$ and $v_t= (v^1_t,v^2_t, \ldots, v^N_t$) the $N$ $d$-dimensional positions and velocities of the particles, the system's evolution is governed by the stochastic dynamics
\begin{equation}
	\label{eqn:vicsek}
	\dot x_t = v_t, \qquad \dot v_t =  f(x_t, v_t) -\gamma v_t  + \eta_{t}.
\end{equation}
Here, $1/\gamma$ is a persistence time, $\eta_t$ is a zero-mean white noise process with covariance $\langle \eta^i_t\eta^j_s\rangle = 2\gamma v_{*}^{2}\delta_{ij}  \delta(t-s) \text{Id}$, and $v_{*}$ is a characteristic velocity.
The active force $f=(f^1,f^2,\ldots, f^N)$ is given by the dynamical XY-like model~\citep{stanley_dependence_1968},
\begin{equation}
	\label{eq:potential}
	f^i(x, v) =  \sum_{j=1}^{N} \big(v^j - v^i\big)K\big(|x^i - x^j|\big),
\end{equation}
where the rotation-invariant kernel $K$ is a smooth variant of the Heaviside step function (SI Appendix).
Physically, the active force $f$ aligns the velocities of the particles in the range of the kernel $K$.
Eq.~\cref{eqn:vicsek} gives rise to a rich and complex dynamics with multiple coexisting flocks that dynamically form and break apart, as shown in Fig.~\ref{fig:overview}.

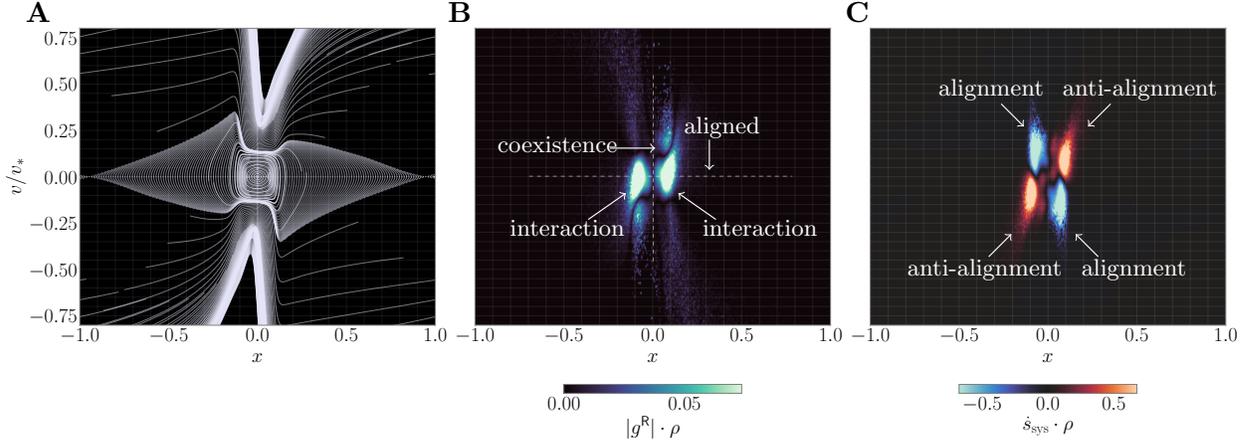
\begin{figure*}[!t]
	\begin{tabular}{c}
		\begin{overpic}[width=\textwidth]{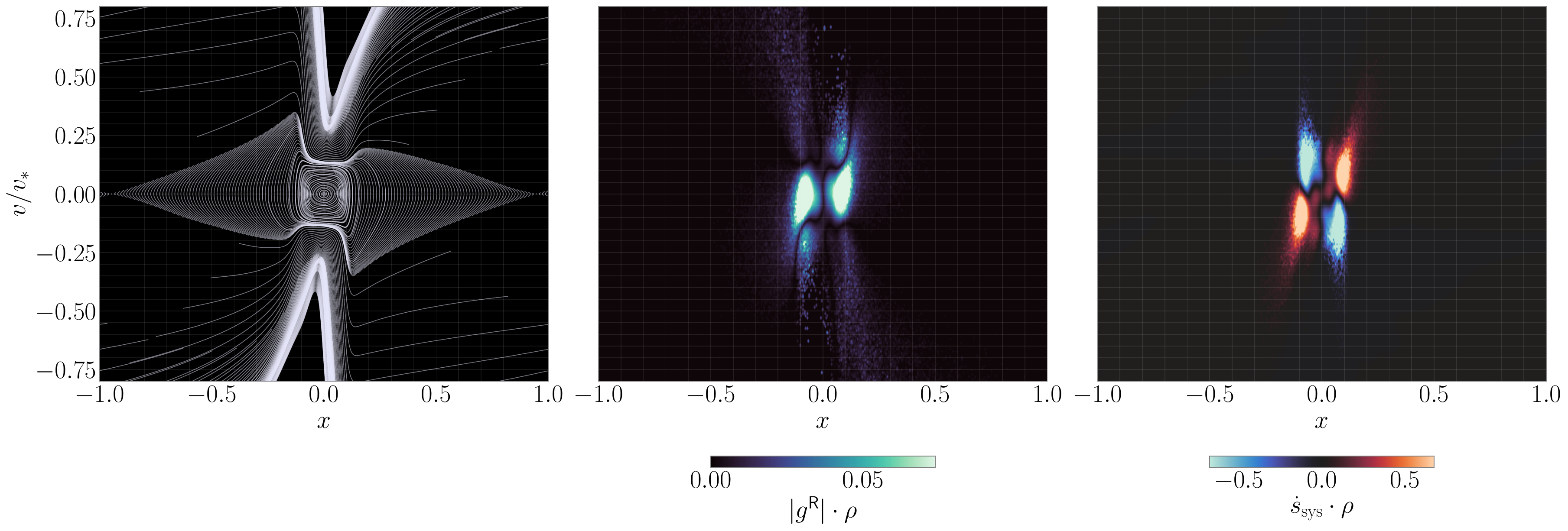}
			\put( 2, 34){\textbf{A}}

			\put(36, 34){\textbf{B}}

			\put(41, 16.75){\footnotesize\textcolor{white}{interaction}}
			\put(47.25,18){\tikz\draw[->][white, line width=0.15mm] (0,0) -- (0.5,0.35);}

			\put(56.5, 16.75){\footnotesize\textcolor{white}{interaction}}
			\put(54.5,18){\tikz\draw[->][white, line width=0.15mm] (0,0) -- (-0.5,0.35);}

			\put(55, 25){\footnotesize\textcolor{white}{aligned}}
			\put(57, 22){\tikz\draw[->][white, line width=0.15mm] (0,0) -- (0,-0.4);}
			\put(42.5,21.5){\tikz\draw[white, opacity=0.5, dash pattern=on 1.5pt off 1.5pt, line width=0.1mm] (0,0) -- (3.5,0);}

			\put(40, 23.5){\footnotesize\textcolor{white}{coexistence}}
			\put(49, 23.75){\tikz\draw[->][white, line width=0.15mm] (0,0) -- (0.6,0);}
			\put(52.5,14.5){\tikz\draw[white, opacity=0.5, dash pattern=on 1.5pt off 1.5pt, line width=0.1mm] (0,0) -- (0,-2.5);}

			\put(68., 34){\textbf{C}}

			\put(75.5, 28){\footnotesize\textcolor{white}{alignment}}
			\put(80.5, 25.5){\tikz\draw[->][white, line width=0.15mm] (0,0) -- (0.3,-0.3);}

			\put(85.5, 28){\footnotesize\textcolor{white}{anti-alignment}}
			\put(87.5, 25.5){\tikz\draw[->][white, line width=0.15mm] (0,0) -- (-0.3,-0.3);}

			\put(73, 13.5){\footnotesize\textcolor{white}{anti-alignment}}
			\put(79.5,15){\tikz\draw[->][white, line width=0.15mm] (0,0) -- (0.3,0.3);}

			\put(87, 13.5){\footnotesize\textcolor{white}{alignment}}
			\put(86.5, 15){\tikz\draw[->][white, line width=0.15mm] (0,0) -- (-0.3,0.3);}

		\end{overpic}
	\end{tabular}
	\caption{%
	\textbf{Two particles: Phase portrait and EPR.}
	(A) Flow lines of the probability current on the NESS.
	Pendulum-like limit cycles of alignment and anti-alignment can be seen.
	Comparison with (B) and (C) highlights alternating regions of $\sdottot \neq 0$ and $\sdottot = 0$, as well as alternating regions of system entropy production ($\sdotsys > 0$) and system entropy consumption ($\sdotsys < 0$) as the particles oscillate in and out of the aligned state.
	(B) Phase space distribution of $\sdottot$, which concentrates during the particle-particle interaction ($|g^{\rev}|$ is visualized rather than $|g^{\rev}|^{2}$ to reduce the influence of outliers).
	(C) Phase space distribution of $\sdotsys$, which concentrates in alternating, asymmetric lobes of entropy production and entropy consumption.
	Entropy production occurs during anti-alignment -- regions where $\sign(x) = \sign(v)$ -- and indicates the annihilation of order.
	Entropy consumption occurs during alignment, which happens when $\sign(x) = -\sign(v)$, and signifies the creation of order.
	}%
	\label{fig:vicsek_lowd_EPR}
\end{figure*}

\paragraph{Time-reversal.}
To quantify the non-equilibrium nature of the dynamics~\cref{eqn:vicsek}, we investigate the structure of its TRS breaking events~\cite{ferretti_signatures_2022,chetrite_fluctuation_2008}.
To this end, given an infinitely-long trajectory $(x_t,v_t)$ evolving over the system's non-equilibrium steady state (NESS), we define the corresponding time-reversed trajectory as $(x^R_t,v^R_t) \equiv (x_{-t}, -v_{-t})$.
We show in the end matter that this time-reversed process evolves according to
\begin{equation}
	\label{eqn:reverse_2}
	\dot{x}_{t}^{\rev} = v_{t}^{\rev}, \quad \dot{v}_{t}^{\rev}  =  f(x_t^\rev, v_t^\rev) -\gamma v_t^{\rev} - 2g^\rev(x_t^\rev, v_t^\rev) + \eta_t.
\end{equation}
Compared to \cref{eqn:vicsek}, the reverse-time equation~\cref{eqn:reverse_2} contains an additional force $g^\rev(x,v) = -g(x,-v) $ with $g$ given by
\begin{equation}
	\label{eq:gdef}
	g(x,v) =  f(x,v) - \gamma v - \gamma v_*^2\nabla_{v} \log \rho(x,v).
\end{equation}
Above, $\rho(x,v)$ is the NESS probability density function, which enters here through the so-called \textit{score function} $\nabla_{v} \log \rho$~\citep{song_score-based_2021, hyvarinen_estimation_2005}.
We find that $g = 0$ in the absence of the active force, while $g\not=0$ when $f\not=0$.
The function $g$ also characterizes the probability current on the NESS, which is a vector field whose local value at $(x,v)$ is precisely $(v,g(x,v))\rho(x,v)$, as discussed in the End Matter.
The discrepancy between the forward and reverse-time dynamics indicates that the active nature of the system breaks TRS, which we aim to quantify by estimating $g$ with deep learning.
We emphasize that this calculation cannot be completed with standard numerical methods because $g(x,v)$ is a high-dimensional function that depends, \textit{a-priori}, on the positions and velocities of \textit{all} particles in the system through the score $\nabla_{v} \log \rho(x,v)$.
We show in the SI Appendix that $g$ can also be expressed as
\begin{equation}
	\label{eqn:pflow_vel}
	\begin{aligned}
		g(x, v) & = \langle \dot{v}_t \: |\: (x_t, v_t) = (x, v)\rangle,
	\end{aligned}
\end{equation}
where $\langle \:\cdot\: |\: (x_t, v_t) = (x, v)\rangle$ denotes an average over the NESS conditioned on the event $(x_t, v_t) = (x, v)$. Below, we will make use of~\eqref{eqn:pflow_vel} to design a variational objective to estimate $g$.

\paragraph{Total entropy production rate.}
One way to measure the global breakdown of TRS is to calculate the total EPR, which is defined as the Kullback-Leibler divergence between the path measure $\calP_t$ of~\eqref{eqn:vicsek} and the path measure $\calP_t^\rev$ of its time-reversal~\cref{eqn:reverse_2} over a horizon $t > 0$~\citep{crooks_entropy_1999,lebowitz_gallavotticohen-type_1999}.
While a useful object, it is well-known that the total EPR does not capture the spatial structure of TRS breaking events because it is a single scalar~\citep{ro_model-free_2022, nardini_entropy_2017, fodor_how_2016}.
To overcome this limitation, we introduce the microscopic phase space quantity,
\begin{equation}
	\label{eqn:sdot_tot_micro}
	\sdottot(x, v) = \left\langle\frac{d}{dt}\log\left(\frac{\calP_{t}(\phi_t)}{\calP_{t}^\rev(\phi_t)}\right) \bigg| (x_t, v_t) = (x, v)\right\rangle.
\end{equation}
In~\eqref{eqn:sdot_tot_micro}, $\phi_{t} =\{x_{\tau},v_{\tau}\}_{\tau\in[0,t]}$ is a path of length $t$ on the NESS and $\langle\:\cdot\: |\:(x_{t}, v_{t}) = (x, v)\rangle$ denotes an average over all such paths of the forward-time dynamics that end at $(x, v)$.
Physically,~\cref{eqn:sdot_tot_micro} measures how much less likely a trajectory of the forward-time dynamics becomes under the reverse-time path measure when its duration is extended infinitesimally.
We show in the SI Appendix the following remarkable relation between $\sdottot(x, v)$ and $g^{\rev}(x, v)$,
\begin{equation}
	\label{eqn:sdot_tot_micro_decomp}
	\begin{aligned}
		\sdottot(x, v) & = \frac{1}{\gamma v_{*}^{2}}|g^\rev(x, v)|^{2} = \frac{1}{\gamma v_*^2}\sum_{i=1}^N |g^{\rev,i}(x, v)|^2,
	\end{aligned}
\end{equation}
which demonstrates the central role of $g^{\rev}$ in measures of TRS breaking.
In~\cref{eqn:sdot_tot_micro_decomp}, we have highlighted that our phase space definition of the total EPR may be decomposed into contributions from individual particles, which will enable us to visualize its spatial distribution.
We show in the SI Appendix
using a path integral approach~\citep{onsager_fluctuations_1953,machlup_fluctuations_1953} that~\cref{eqn:sdot_tot_micro_decomp} recovers the macroscopic total EPR upon averaging over the NESS, i.e.
\begin{equation}
	\label{eqn:sdot_tot:0}
	\Sdottot = \lim_{t\rightarrow\infty}\frac{1}{t}\kl{\calP_{t}}{\calP^{\rev}_{t}} = \langle \sdottot(x, v)\rangle,
\end{equation}
so that $\sdottot$ is a natural microscopic counterpart to $\Sdottot$.

\paragraph{System entropy production rate.}
The stochastic entropy of the system~\citep{seifert_entropy_2005,seifert_stochastic_2012} is an information-theoretic quantity defined as the negative logarithm of the probability density function $\rho$ for the NESS evaluated along a trajectory $(x_t, v_t)$ of~\eqref{eqn:vicsek},
\begin{equation}
	\label{eqn:system_entropy}
	s_{\text{sys}}^{\text{stoch}}(t) = -\log\rho(x_{t}, v_{t}),
\end{equation}
so that its average over the NESS gives the Gibbs entropy.
Similar to~\cref{eqn:sdot_tot_micro}, we define the corresponding system EPR as the conditional average of its time derivative~\citep{boffi_deep_2024}
\begin{equation}
	\label{eqn:sdot_sys}
	\sdotsys(x, v) = - \left\langle\frac{d}{dt}\log\rho(x_{t}, v_{t}) \:|\: (x_{t}, v_{t}) = (x, v)\right\rangle,
\end{equation}
which converts the time-dependent quantity~\cref{eqn:system_entropy} into a phase space quantity.
As we show in the end matter, leveraging that $\rho$ solves a stationary Fokker-Planck equation yields a connection between the system EPR and the divergence of $g$,
\begin{equation}
	\label{eqn:sdot_sys_expression}
	\sdotsys(x, v) = \nabla_{v} \cdot g(x, v) = \sum_{i=1}^{N} \nabla_{v^{i}}\cdot g^{i}(x, v).
\end{equation}
Again, we have highlighted that $\sdotsys(x, g)$ may be broken down into contributions from individual particles.

\paragraph{Estimating the current velocity.}
The expressions~\cref{eqn:sdot_tot_micro} and~\cref{eqn:sdot_sys_expression} highlight the fundamental relation between $g$ and the measures $\sdottot$ and $\sdotsys$ of TRS breaking.
We show in the SI Appendix that $g$ is the unique minimizer over $\hat{g}$ of the variational objective
\begin{equation}
	\label{eqn:strato_loss}
	\mathcal{L}[\hat{g}] = \frac{1}{T}\E\left[\int_0^T |\hat{g}(x_t, v_t)|^2dt - 2\hat{g}(x_t, v_t)\circ dv_t\right],
\end{equation}
where $\circ$ denotes a Stratonovich product and where $T > 0$ is an arbitrary time horizon.
We estimate $\hat{g}$ by training a neural network to minimize~\eqref{eqn:strato_loss} using the Adam optimizer~\citep{kingma_adam_2017}.
As described in the end matter, we use a graph neural network~\citep{battaglia_relational_2018} that enforces two key physical symmetries in the system: permutation equivariance amongst the particles and translation invariance.

\begin{figure*}[!th]
	\centering
	\begin{tabular}{c}
		\begin{overpic}[width=\textwidth]{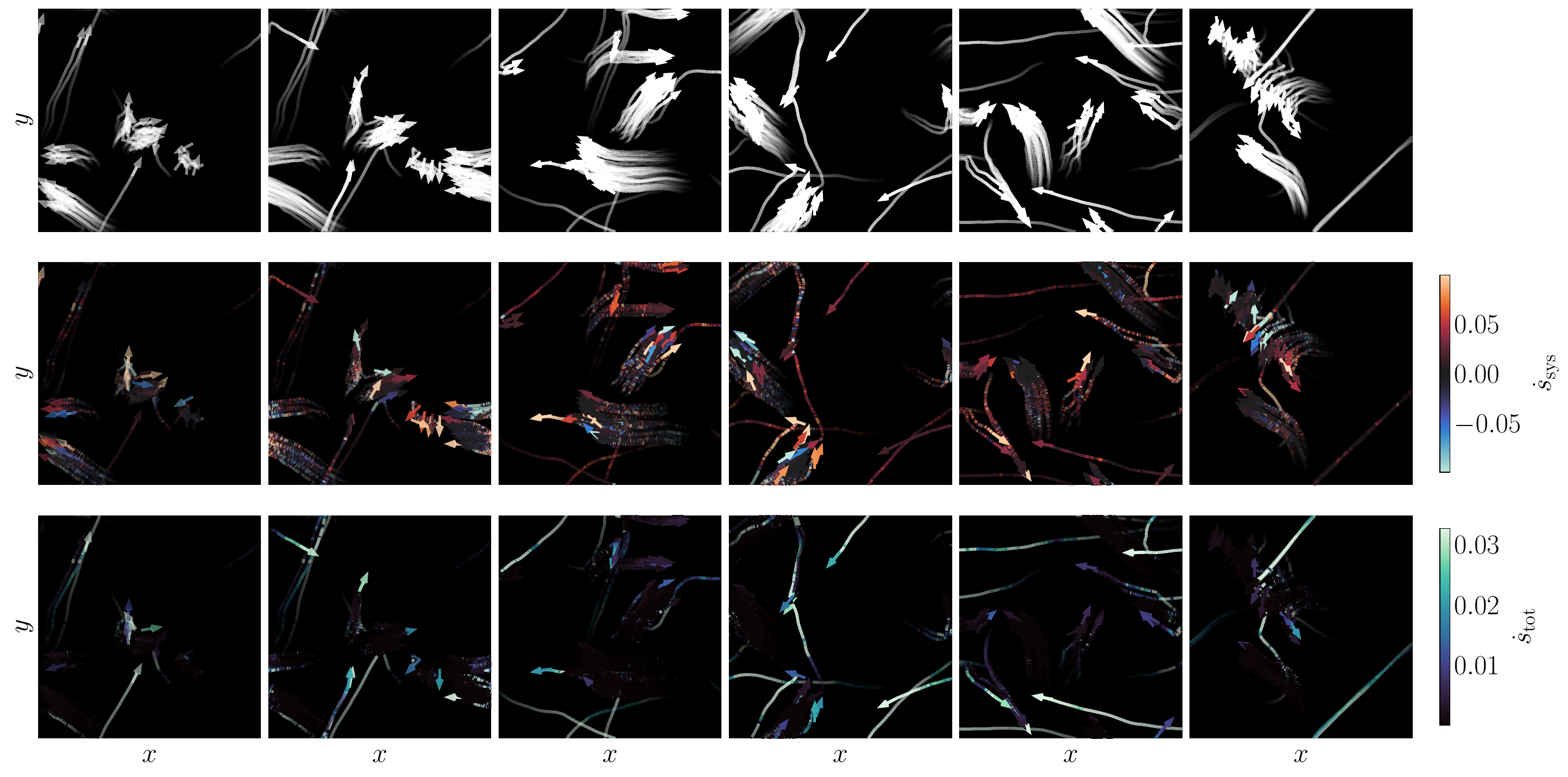}
			\put(-1, 51.0){\textbf{A}}
			\put(-1, 34.0){\textbf{B}}
			\put(-1, 17.0){\textbf{C}}
		\end{overpic} \\

		\hspace{-12.5mm}
		\begin{tabular}{cc}
			\centering
			\begin{overpic}[width=0.48\textwidth]{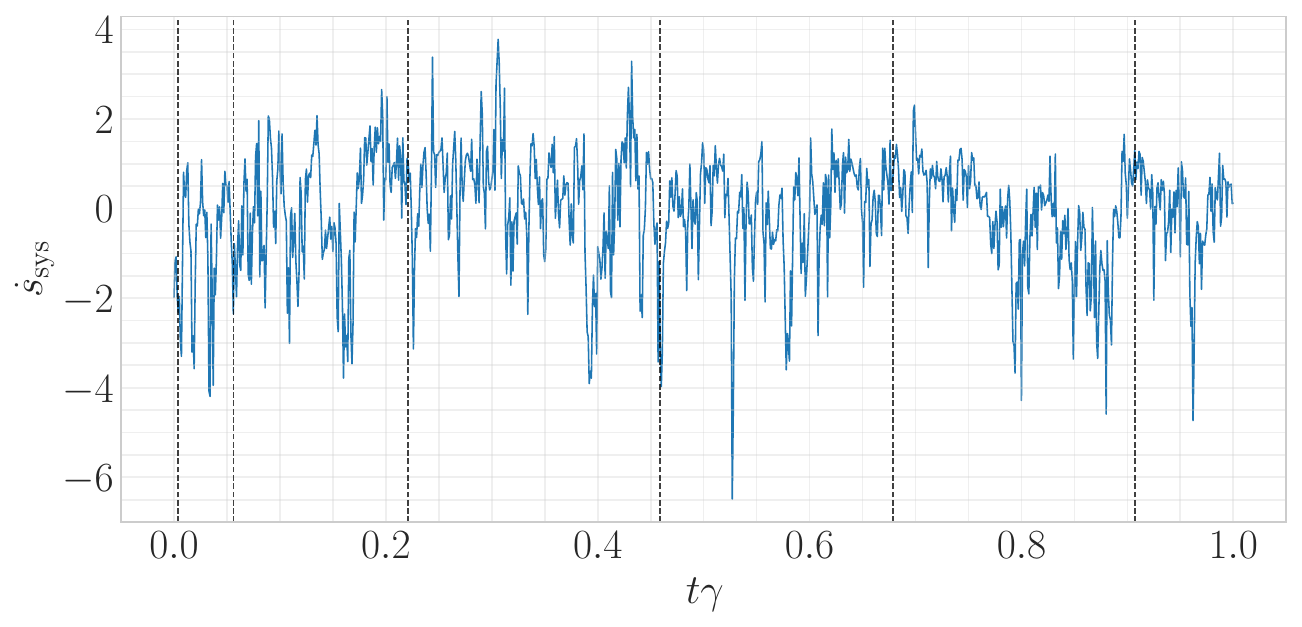}
				\put(1.5, 47.5){\textbf{E}}
			\end{overpic} &
			\begin{overpic}[width=0.48\textwidth]{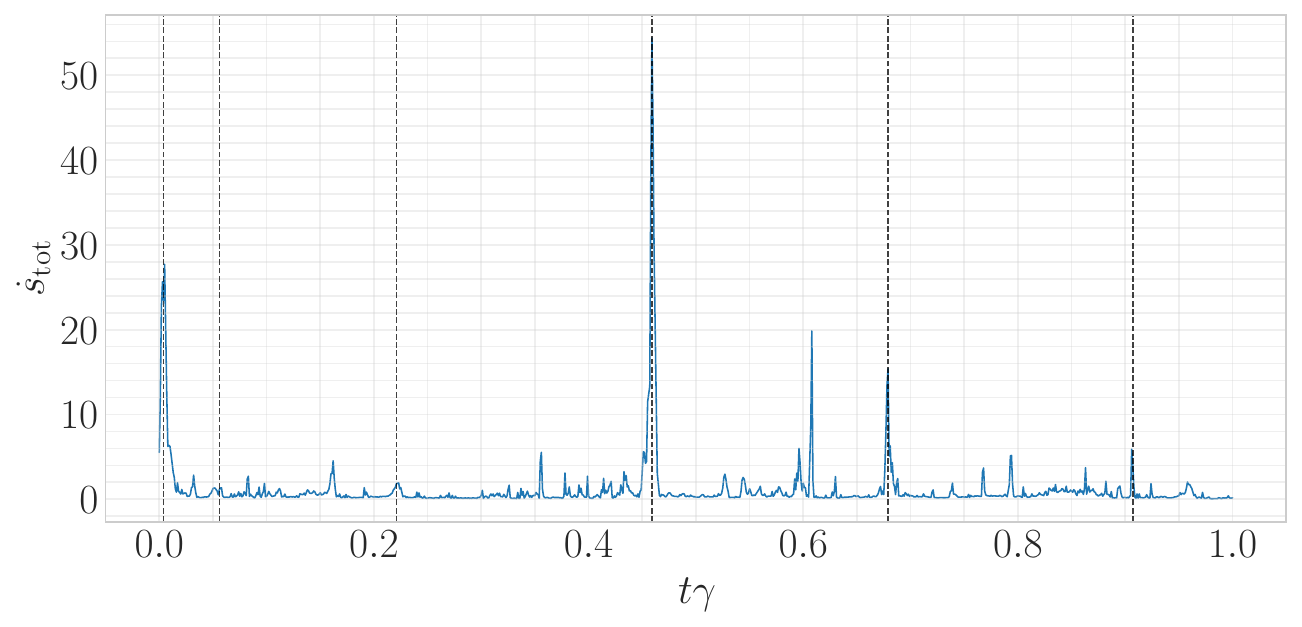}
				\put(1.5, 47.5){\textbf{F}}
			\end{overpic}
		\end{tabular}
	\end{tabular}
	\caption{\textbf{Sixty-four particles in two dimensions.}
		The reader is strongly encouraged to view the accompanying movie \href{https://www.dropbox.com/scl/fi/edh2nuzxgozd4z1wxby2e/sde_entropy_movie_plot.mp4?rlkey=fh156967p9msfp6fg85ktpq41&st=ljvqq6yo&dl=0}{here}, as well as the longer movie \href{https://www.dropbox.com/scl/fi/wttsp4rhpb6kgfiwesnxi/sde_entropy_movie_plot_long.mp4?rlkey=9chq2pru34csiz0joonj0g57d&st=ywv2k0lw&dl=0}{here}.
		(A) Uncolored reference depiction of the particle trajectories.
		Frames chosen based on large spikes in $\sdottot$, shown by vertical lines in (E) and (F).
		(B) Particle contributions to $\sdotsys$.
		Particles exhibit negative system EPR during alignment and positive system EPR during anti-alignment.
		Positive values occur on the boundary of flocks, while negative values occur both on the boundary and during collisions.
		(C) Particle contributions to $\sdottot$.
		The signal is similar to $\sdotsys$, but is more dominated by outliers, and does not display signed information about the creation or annihilation of order.
		Particles primarily contribute to $\sdottot$ during collisions and during flock breakup.
		(E) Time series of $\sdotsys(x_t, v_t)$.
		Large negative spikes indicate an increase in system order, such as during a merger between two flocks.
		Positive spikes indicate a decrease in system order, corresponding to particles leaving alignment, as driven by random fluctuation on flock boundaries.
		(F) Time series of $\sdottot(x_t, v_t)$.
		Large spikes typically correspond to flock breakup or flock formation.
	}%
	\label{fig:vicsek_N64_EPR}
\end{figure*}

%% file: results.tex
\section{Results}
\paragraph{Low-dimensional system.}
To build physical intuition, we first consider the case of $N=2$ particles in $d=1$ dimension.
By transformation to displacement coordinates $x= x^{1} - x^{2}$ and $v = v^{1} - v^{2}$, the system admits an equivalent two-dimensional description that enables us to visualize the EPR over the entire phase space (\cref{fig:vicsek_lowd_EPR}).
To remove the influence of low-probability outliers, all quantities are weighted by the stationary density $\rho$.

Flow lines of the probability current $(v, g(x,v))\rho(x,v)$ on the NESS are shown in~\cref{fig:vicsek_lowd_EPR}A: the phase portrait displays limit cycles that correspond to successive alignment and anti-alignment of the particles.
For large values of $|v|$, the particles do not always enter a bound state, and instead deflect off of each other.

An $(x, v) \mapsto (-x, -v)$ symmetry is clear in all panes, which corresponds to permutation symmetry between the two particles in this frame of reference.
The horizontal line $v = 0$ corresponds to the ordered, aligned phase, while the vertical line $x = 0$ corresponds to spatial superposition.
Both the total EPR (\cref{fig:vicsek_lowd_EPR}B) and the system EPR (\cref{fig:vicsek_lowd_EPR}C) are nonzero when the particles interact, corresponding to the region $|x| \lesssim 0.25$ and $|v| \gtrsim 0$.
Moreover, both vanish around the aligned phase $|v| \approx 0$, $|x| \approx 0$ and in the disordered gas $|x| \gtrsim 0.25$.

The signal in both $\sdottot$ and $\sdotsys$ is similar, though $\sdottot$ has lost the sign information contained in $\sdotsys$.
Physically, the sign of $\sdotsys$ indicates the creation or annihilation of order: during anti-alignment and transition towards the disordered gaseous phase, the system EPR is positive, while during alignment and formation of the ordered phase, the system EPR is negative.
These transitions are not captured by the total EPR $\sdottot$, which only detects the particle-particle interaction.
Comparison with~\cref{fig:vicsek_lowd_EPR}A confirms a physical picture in which entropy is cyclically consumed and produced as the particles progress between the aligned and disordered states in a periodic pattern.

The phase space depiction of $\sdotsys$ reveals that entropy is produced asymmetrically from how it is consumed, a phenomenon that is not seen in $\sdottot$.
Entropy production occurs over a larger region of phase space, but by the stationarity condition $\partial_{t}\int \log\rho(x, v)\rho(x, v) = 0$, the thermodynamic average of the system EPR must vanish.
To attain zero average, the distribution of entropy consumption is heavy-tailed (\cref{fig:lowd_vicsek_statistics}, SI Appendix), which compensates for its lower phase space volume.

The origin of this heavy tail can be understood with a simple physical picture of the dynamics.
Because the domain has periodic boundary conditions, the two particles will necessarily collide.
At the collision, the interaction causes the particles to align, which produces order and induces a negative spike in $\sdotsys$.
These collision events can lead to abnormally large negative values of $\sdotsys$, thereby producing a heavy tail.
By contrast, anti-alignment and its associated positive values of $\sdotsys$, can only occur via the random Gaussian fluctuations in~\cref{eqn:vicsek}.
These Gaussian statistics ensure that events of large positive $\sdotsys$ are rare.

\paragraph{High-dimensional system.}
We now consider the case of $N=64$ particles in $d=2$ dimensions.
Unlike the two-particle system just considered, here there is no mapping to an equivalent low-dimensional system, and the dimensionality of the phase space is $2Nd = 256$.
This necessitates the use of machine learning, as $g$ is a function defined on the high-dimensional phase space, and is therefore challenging to approximate with traditional discretization or basis function expansion-based techniques.

Results are shown in~\cref{fig:vicsek_N64_EPR}, which depicts time lapses of the dynamics colored by the particle contributions to $\sdottot$ and $\sdotsys$ according to the decompositions given in~\cref{eqn:sdot_tot_micro_decomp,eqn:sdot_sys_expression}.
To gain insight into the global system behavior, time series $\sdotsys(x_t, v_t)$ and $\sdottot(x_t, v_t)$ are shown below; black dashed lines indicate the central frame of each time lapse, chosen adaptively based on events of large $\sdottot$.
An accompanying movie matching the frames is available \href{https://www.dropbox.com/scl/fi/edh2nuzxgozd4z1wxby2e/sde_entropy_movie_plot.mp4?rlkey=fh156967p9msfp6fg85ktpq41&st=ljvqq6yo&dl=0}{at this link}, along with a movie of a longer trajectory \href{https://www.dropbox.com/scl/fi/wttsp4rhpb6kgfiwesnxi/sde_entropy_movie_plot_long.mp4?rlkey=9chq2pru34csiz0joonj0g57d&st=ywv2k0lw&dl=0}{here}.
The reader is highly encouraged to view the movies, as they provide more physical insight into the dynamics than can be obtained by viewing still frames.

The time series $\sdottot(x_t, v_t)$ and $\sdotsys(x_t, v_t)$ are erratic, and display frequent but irregular spikes.
Spikes in $\sdottot(x_t, v_t)$ correspond primarily to flock breakup, during which one or more particles are driven out of their flock by random noise.
More informatively, $\sdotsys(x_t, v_t)$ displays consistent fluctuations around zero, which are reflective of the constant alignment and anti-alignment of particles on flock boundaries.
These particles are driven towards disorder by noise and towards order by the alignment interaction, the balance of which creates a fluctuating trajectory for the EPR.
Deep in the flock center, where a large number of alignment interactions preserves the direction of flight, particles maintain $\sdotsys^{i}(x_t, v_t) = 0$.
Large negative spikes correspond to collisions between existing flocks, which leads to the formation of a single larger flock and drives an associated increase in order.
Similarly, large positive spikes correspond to the breakup of flocks.
These observations confirm the physical picture put forth by the simple two-particle model system: $\sdotsys$ contains signed information about the creation and annihilation of order.
By contrast, $\sdottot$ is less sensitive, and mostly indicates the presence and magnitude of interactions between particles.

\begin{figure}[!t]
	\centering
	\begin{overpic}[width=0.7\textwidth]{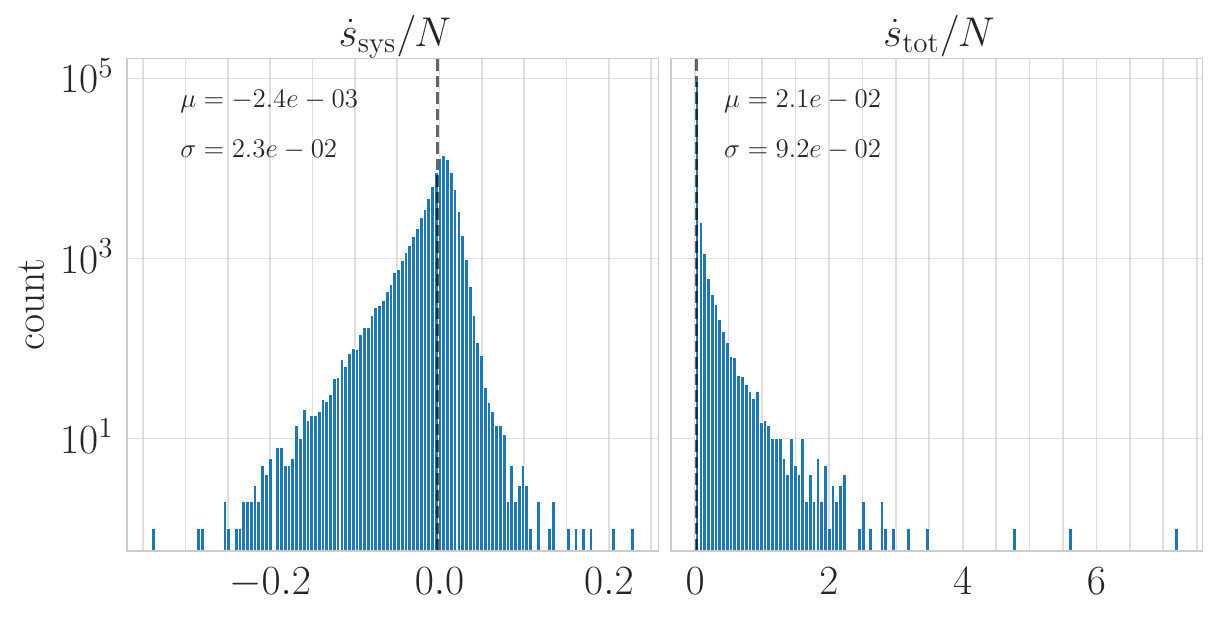}
		\put( 7.5, 48){\textbf{A}}
		\put(55.5, 48){\textbf{B}}
	\end{overpic}
	\caption{
		\textbf{EPR statistics}.
		(A) Density of the per-particle system EPR. The distribution exhibits asymmetry between entropy production and entropy consumption.
		(B) Density of the per-particle total EPR. The distribution has a heavy tail, but does not distinguish between production and consumption. 
	}
	\label{fig:N64_vicsek_statistics}
\end{figure}

Distributions of the average per-particle $\sdotsys$ and $\sdottot$ on the NESS $\rho$ are shown in~\cref{fig:N64_vicsek_statistics}, where exponential tails are seen in both cases.
Similar to the low-dimensional setting discussed earlier, the distribution of $\sdotsys$ is asymmetric: a heavier tail of entropy consumption is balanced by more frequent events of lower-magnitude entropy production.
This asymmetry can again be rationalized by a simple physical picture.
Instances of large entropy consumption are driven by flock merger events that occur due to the periodic boundary conditions.
Entropy is consistently produced in small amounts as random fluctuations cause particles on flock boundaries to peel away towards the free particle state.

\begin{figure}[!t]
	\centering
	\begin{overpic}[width=0.7\textwidth]{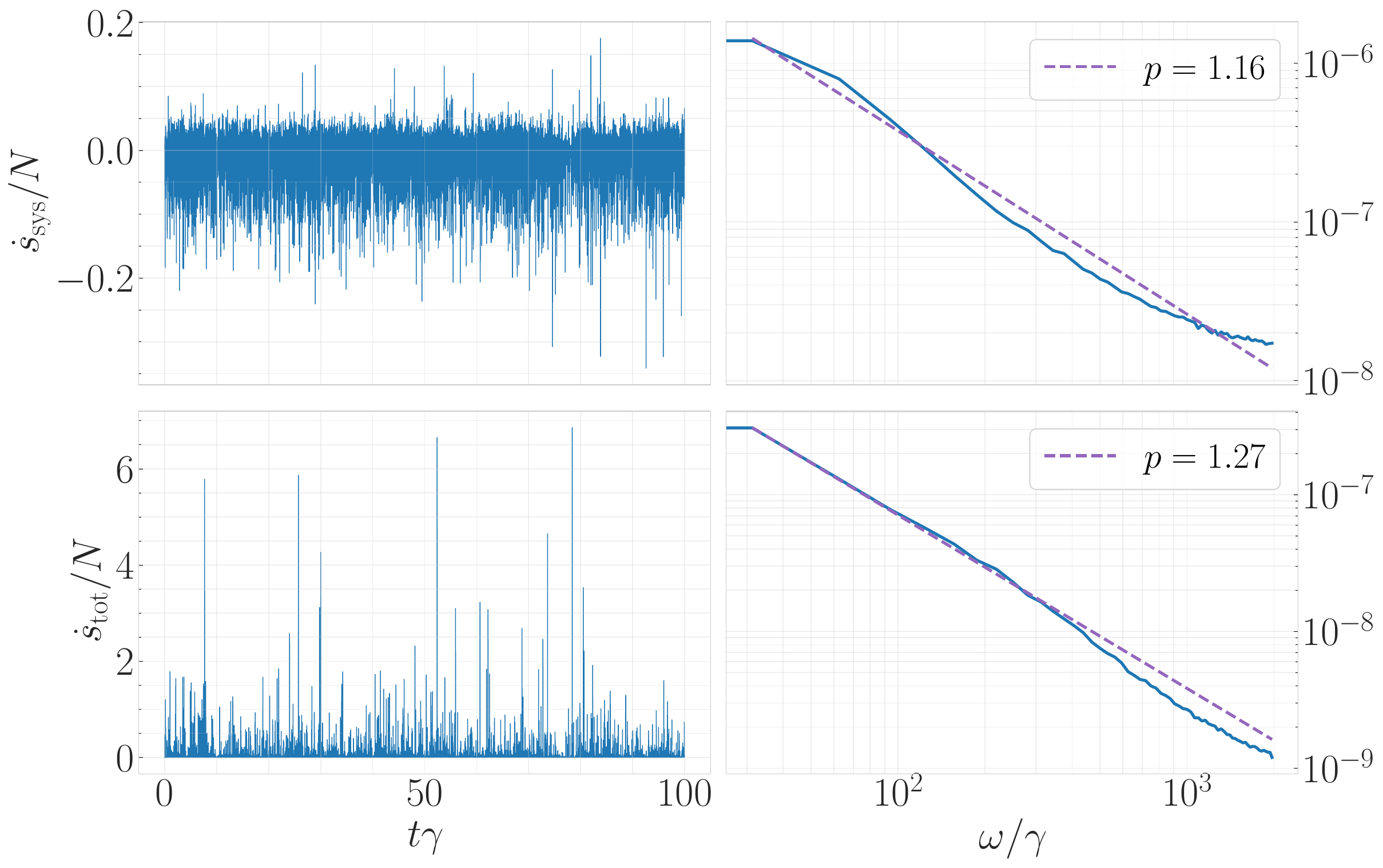}
		\put( 2.5, 62.5){\textbf{A}}
		\put( 2.5,   30){\textbf{B}}
	\end{overpic}
	\caption{
		\textbf{Power spectral density}.
		Time series and PSD for the system EPR (A) and the total EPR (B).
		Dashed line on PSD indicates nonlinear fit to $c/\omega^{p}$.
	}
	\label{fig:N64_vicsek_psd}
\end{figure}

In~\cref{fig:N64_vicsek_psd}, we quantify the intermittent dynamics of $\sdotsys(x_t, v_t)$ and $\sdottot(x_t, v_t)$ by estimating the power spectral density (PSD) of long, stationary time series.
In both cases we identify the appearance of pink noise with a power law exponent that is lower than the expected exponent of $2$ for pure Brownian motion.
A power law PSD is expected for intermittent time series~\citep{manneville_intermittency_1980,hirsch_theory_1982} and has been identified in flocking models in earlier work~\citep{huepe_intermittency_2004}, but has not been observed in the EPR due to the difficulties associated with its computation.

%% file: conc.tex
\section{Conclusion}
In this work, we introduced a model-free machine learning approach to compute the system and total entropy production rates of nonequilibrium dynamical systems.
We found that the system EPR contains signed information illustrating the creation and annihilation of order as particles enter or leave the aligned state, while the total EPR cannot distinguish between these two thermodynamic phenomena.
In both cases, we observed intermittent dynamics dominated by collisions between flocks, which we quantified by identifying heavy-tailed distributions in the EPR and $1/f$ noise in the associated power spectral densities.
Future work will investigate the structure of the EPR in driven systems subject to an external control protocol and in equilibrium systems as they relax towards the Gibbs density from an initially non-stationary state.
In addition, we plan to investigate the realistic experimental setting where only a subset of dynamical variables can be observed, and to understand how the choice of these variables affects the resulting predictions for the EPR.

%% file: end_matter.tex
\section{End matter}
\paragraph{Time reversal.}
\noindent
We now derive the time-reversed dynamics given in~\cref{eqn:reverse_2}.
Complete details of the steps shown here are provided in the SI Appendix.
To proceed, it is convenient to consider the dynamics governing the evolution of the system's time-dependent microscopic probability density function $\rho_t(x,v)$ towards stationarity.
This quantity satisfies
\begin{equation}
   \label{eqn:td_fpe} 
   \begin{aligned}
   \partial_t \rho_t + \nabla_x \cdot\left(v \rho_t\right) + \nabla_v \cdot\left([f-\gamma v]\rho_t\right) &= \gamma v_*^2\Delta_v \rho_t,
   \end{aligned}
\end{equation}
solved forwards in time on $t\in[0,\infty)$.
By construction, the NESS density is $\rho = \lim_{t\rightarrow\infty} \rho_t$ for any initial $\rho_0=\rho_{\text{init}}$.
We define the time-reversed PDF by the relation
\begin{equation}
    \label{eqn:rho_rev_def}
     \rho_t^\rev(x, -v) = \rho_{-t}(x, v).
\end{equation}
The definition in~\cref{eqn:rho_rev_def} ensures that particles continue to move in the direction of their velocity in the time-reversed dynamics.
From~\cref{eqn:rho_rev_def}, we have that $\partial_t \rho_t^\rev(x, v) = -\partial_t \rho_{-t}(x, -v)$, so
\begin{equation}
    \label{eqn:dt_rhot_rev_steps_1}
    \begin{aligned}
        \partial_t \rho_t^\rev &= -\nabla_x \cdot\left(v \rho_{t}^\rev\right) + \nabla_v \cdot\left([f-\gamma v]\rho_{t}^\rev\right) - \gamma v_*^2\Delta_v \rho_{t}^\rev,
    \end{aligned}
\end{equation}
where all quantities are evaluated at $(x, v)$ and we used $f(x,-v)=-f(x,v)$.
Introducing 
\begin{equation}
    \label{eq:gRdef:em}
    g_t^\rev(x,v) =  f(x,v) - \gamma v - \gamma v_*^2\nabla \log \rho^\rev_t(x,v),
\end{equation}
we may then write~\eqref{eqn:dt_rhot_rev_steps_1} as
\begin{equation}
    \label{eqn:dt_rhot_rev_steps_3}
    \begin{aligned}
        \partial_t \rho_t^\rev &= -\nabla_x \cdot\left(v \rho_{t}^\rev\right) - \nabla_v \cdot\left([f-\gamma v -2g^\rev_t ]\rho_{t}^\rev\right) + \gamma v_*^2\Delta_v \rho_{t}^\rev,
    \end{aligned}
\end{equation}
which is the  Fokker-Planck equation for the stochastic dynamics
\begin{equation}
    \label{eq:reverse:sde:em}
    \begin{aligned}
    \dot{x}_t^\rev &= v_t^\rev,\\
    \dot{v}_t^\rev &= f(x_t^\rev, v_t^\rev) - \gamma v_t^\rev - 2g_t^\rev(x_t^\rev, v_t^\rev) + \eta_t.
    \end{aligned}
\end{equation}
Eq.~\eqref{eq:gRdef:em} implies that  $g^\rev_{-t}(x,-v) = g_t(x,v) $ with $g_t$ given by
\begin{equation}
    \label{eq:gdef:em}
    g_t(x,v) =  f(x,v) - \gamma v - \gamma v_*^2\nabla \log \rho_t(x,v).
\end{equation}
At stationarity we then have $g_t = g =  f - \gamma v - \gamma v_*^2\nabla \log \rho$, and we recover~\eqref{eq:gdef}.
We show in the SI Appendix that this expression is equivalent to $g(x,v) = \langle \dot v_t | (x_t,v_t) = (x,v)\rangle$.

\paragraph{Transport equation and probability flow.}
The NESS density $\rho$ for~\eqref{eqn:vicsek} is the solution of the stationary many-body Fokker-Planck equation (FPE)
\begin{equation}
	\label{eqn:fpe}
	\begin{aligned}
	  \nabla_{x}\cdot\left(v\rho\right) + \nabla_{v} \cdot \left((f - \gamma v)\rho\right) = \gamma v_{*}^{2}\Delta_{v}\rho.
	\end{aligned}
\end{equation}
By definition of $g$ in~\cref{eqn:pflow_vel}, this FPE may be re-written as the stationary transport equation
\begin{equation}
	\label{eqn:transport}
	\begin{aligned}
	  \nabla_{x}\cdot\left(v\rho\right) + \nabla_{v} \cdot \left(g\rho\right) = 0.
	\end{aligned}
\end{equation}
which indicates that the probability current on the NESS is the vector field $(v,g(x,v))\rho(x,v)$.
The flow lines of this probability current define the probability flow~\citep{boffi_probability_2023,maoutsa_interacting_2020,song_score-based_2021}, i.e., the phase-space curves $(x^c(t), v^c(t))_{t\in \R}$ such that
\begin{equation}
\label{eqn:pflow}
\begin{aligned}
    \dot{x}^c_t &= v^c_t,\\
    \dot{v}^c_t &= g(x^c_t, v^c_t).
\end{aligned}
\end{equation}
The curves defined by~\cref{eqn:pflow} satisfy that if $(x^c_0, v^c_0) \sim \rho$, then $(x^c_t, v^c_t) \sim \rho$ for all $t\in \R$; i.e., similar to the stochastic dynamics~\cref{eqn:vicsek}, the deterministic dynamics~\cref{eqn:pflow} preserves the NESS.

\paragraph{Estimating the loss.} 
Let $i = 1, \hdots, N$ denote the particle index, $\alpha = 1, \hdots, n$ denote an index over trajectories, and let $(x_{t}^{i, \alpha}, v_{t}^{i, \alpha})$ for $t \in [0, T]$ denote a dataset of trajectories.
Because we consider the estimation of probability flows on a nonequilibrium steady state, $T$ is arbitrary. 
We may therefore take $T = \Delta t$ for a fixed timestep $\Delta t > 0$ and approximate $\calL \approx \hat \calL$ as the empirical mean
\begin{equation}
	\label{eqn:strato_loss_disc}
	\begin{aligned}
      &\hat{\mathcal{L}}[\hat{g}]\\
      &= \frac{1}{n N}\sum_{\alpha=1}^{n}\sum_{i=1}^{N}|\hat{g}^{i}(x_{t}^{\alpha}, v_{t}^{\alpha})|^{2}\Delta t - \hat{g}^{i}(x_{t+\Delta t}^{\alpha}, v_{t+\Delta t}^{\alpha})\cdot(v_{t+\Delta t}^{i, \alpha} - v_{t}^{i, \alpha})- \hat{g}^{i}(x_{t}^{\alpha}, v_{t}^{\alpha})\cdot(v_{t+\Delta t}^{i, \alpha} - v_{t}^{i, \alpha}).
	\end{aligned}
\end{equation}
In practice, we minimize the discrete approximation~\cref{eqn:strato_loss_disc} after generating $n$ samples from $\rho$ by simulating the equation~\cref{eqn:vicsek}.
The single timestep required for~\cref{eqn:strato_loss_disc} can be generated during loss evaluation.

\paragraph{Network architecture.}
We make use of a graph neural network architecture
\begin{equation}
	\label{eqn:gnn}
	\begin{aligned}
		\hat{g}^{i}(x, v) = \psi\left(\sum_{j=1}^{N}\phi\left(x^{i} - x^{j}, v^{i}, v^{j}\right)\right),
	\end{aligned}
\end{equation}
where $\phi$ and $\psi$ are a learnable encoder and decoder pair that we represent with fully-connected networks.
Due to the sum pooling of the encoder states,~\cref{eqn:gnn} enforces permutation symmetry.
Due to its dependence only on differences in positions, the architecture enforces translation invariance.
We take $\phi$ and $\psi$ to be four layer networks with $2048$ neurons per layer.

%% file: app.tex
\section{Time reversal derivation}
\noindent
We now derive the time-reversed dynamics given in~\cref{eqn:reverse_2} in greater detail than shown in the End Matter.
Recall that we have the time-dependent Fokker-Planck equation governing the evolution towards stationarity,
\begin{equation}
	\label{eqn:app:td_fpe}
	\begin{aligned}
		\partial_t \rho_t + \nabla_x \cdot\left(v \rho_t\right) + \nabla_v \cdot\left(b_v\rho_t\right) & = \gamma v_0^2\Delta_v \rho_t, \\
		\rho_0                                                                                         & = \rho_{\text{init}},
	\end{aligned}
\end{equation}
where $b_{v}(x, v) = f(x, v) - \gamma v$.
Further recall that we define the time-reversed density by the relation
\begin{equation}
	\label{eqn:app:rho_rev_def}
	\begin{aligned}
		\rho_t^\rev(x, -v)                           & = \rho_{-t}(x, v), \:\: t\in (-\infty, 0], \\
		\lim_{t\rightarrow-\infty}\rho_t^\rev(x, -v) & = \rho(x, v).
	\end{aligned}
\end{equation}
In the following, we denote by $\partial_{t}$ the derivative with respect to the first argument, $\nabla_{x}$ the gradient with respect to the second, and $\nabla_{v}$ the gradient with respect to the third.
That is, for a function $h_{t}(x, v)$ of $(t, x, v)$, by $\partial_{t} h_{-t}(x, -v)$ we denote evaluation at $(-t, x, -v)$ after computing the derivative.
From~\cref{eqn:app:rho_rev_def}, we have that
\begin{equation}
	\label{eqn:app:dt_rhot_rev_steps_1}
	\begin{aligned}
		\partial_t \rho_t^\rev(x, v) & = -\partial_t \rho_{-t}(x, -v),                                                                                                                                                 \\
		                             & = \nabla_x \cdot\left(-v \rho_{-t}(x,-v)\right) - \nabla_v \cdot\left(\left(f(x, -v) + \gamma v\right)\rho_{-t}(x, -v)\right) - \gamma v_*^2\Delta_v \rho_{-t}(x, -v),          \\
		                             & = \nabla_x \cdot\left(-v \rho_{t}^\rev(x,v)\right) - \nabla_v \cdot\left(\left(f(x, -v) + \gamma v\right)\rho_{t}^\rev(x, v)\right) - \gamma v_*^2\Delta_v \rho_{t}^\rev(x, v), \\
		                             & = \nabla_x \cdot\left(-v \rho_{t}^\rev(x,v)\right) + \nabla_v \cdot\left(\left(f(x, v) - \gamma v\right)\rho_{t}^\rev(x, v)\right) - \gamma v_*^2\Delta_v \rho_{t}^\rev(x, v).
	\end{aligned}
\end{equation}
Above, we have used that $f(x, -v) = -f(x, v)$, as follows from~\cref{eq:potential}.
The final line of~\cref{eqn:app:dt_rhot_rev_steps_1} implies that
\begin{equation}
	\label{eqn:app:dt_rhot_rev_bwd}
	\begin{aligned}
		\partial_t \rho_t^\rev(x, v) + \nabla_x \cdot \left(v \rho_t^\rev(x, v)\right) - \nabla_v \cdot \left(\left(f(x, v) - \gamma v\right)\rho_t^\rev(x, v)\right) = -\gamma v_*^2 \Delta_v \rho_t^\rev(x, v).
	\end{aligned}
\end{equation}
Applying the following relation in terms of the score,
\begin{equation}
	\label{eqn:score_identity}
	\nabla_v \cdot \left(\nabla_v\log\rho_t^\rev(x, v) \rho_t^\rev(x, v)\right) = \Delta_v \rho_t^\rev(x, v),
\end{equation}
we can write the final line of~\cref{eqn:app:dt_rhot_rev_steps_1} as
\begin{equation}
	\label{eqn:app:dt_rhot_rev_steps_2}
	\begin{aligned}
		 & \partial_t \rho_t^\rev(x, v) + \nabla_x \cdot \left(v \rho_t^\rev(x, v)\right) - \nabla_v\cdot \left(\left(f(x, v) - \gamma v - 2\gamma v_0^2\nabla_v\log\rho_t^\rev(x, v)\right)\rho_t^\rev(x, v)\right) \\
		 & \qquad = \gamma v_*^2 \Delta_v \rho_t^\rev(x, v).
	\end{aligned}
\end{equation}
Now, defining
\begin{equation}
	\label{eqn:g_score}
	g_t^\rev(x, v) = f(x, v) - \gamma v - \gamma v_*^2 \nabla_v\log\rho_t^\rev(x, v),
\end{equation}
we may write
\begin{equation}
	\label{eqn:app:dt_rhot_rev_steps_3}
	\begin{aligned}
		\partial_t \rho_t^\rev(x, v) + \nabla_x \cdot \left(v \rho_t^\rev(x, v)\right) + \nabla_v \cdot\left(\left(f(x, v) - \gamma v - 2g_t^\rev(x, v)\right)\rho_t^\rev(x, v)\right) = \gamma v_*^2 \Delta_v \rho_t^\rev(x, v).
	\end{aligned}
\end{equation}
We may then observe the identities
\begin{equation}
	\label{eqn:app:v_symmetry}
	\begin{aligned}
		b_v(x, v)                     & = -b_v(x, -v),                \\
		\nabla_v\log\rho_t^\rev(x, v) & = -\nabla_v\log\rho_t(x, -v).
	\end{aligned}
\end{equation}
Together, the equations in~\cref{eqn:app:v_symmetry} imply the identity
\begin{equation}
	\label{eqn:g_gr}
	g_t^\rev(x, v) = -g_{-t}(x, -v).
\end{equation}
The (forward-time) Fokker-Planck equation~\cref{eqn:app:dt_rhot_rev_steps_2} then admits the associated stochastic dynamics
\begin{equation}
	\begin{aligned}
		\dot{x}_t^\rev & = v_t^\rev,                                                                         \\
		\dot{v}_t^\rev & = f(x_t^\rev, v_t^\rev) - \gamma v_t^\rev - 2g^\rev_t(x_t^\rev, v_t^\rev) + \eta_t,
	\end{aligned}
\end{equation}
At stationarity, $\rho_t(x,v) =\rho^\rev_t(x,-v) = \rho(x,v)$ and $g_t(x,v) = g_{-t}^\rev(x,-v) = g(x,v) = -\gamma v + f(x,v) - \gamma v_*^2 \nabla _v \log\rho(x,v)$.

\section{Formal proof that $g = \langle\dot{v} | x, v\rangle$.}
\noindent
We first show that, given any test function $h(x,v)$, we have
\begin{equation}
	\label{eq:loss:app:2}
	\begin{aligned}
		\lim_{\tau \to 0^+} \frac1{2\tau} \big\langle h(x_{t+\tau},v_{t+\tau})- h(x_{t-\tau},v_{t-\tau}) \big| (x_t,v_t)=(x,v)\big\rangle = v \cdot \nabla _x h(x,v) + g(x,v) \cdot \nabla _v  h(x,v).
	\end{aligned}
\end{equation}
To see this, observe that by stationarity and from the forward and backward SDEs we may write the expectation as
\begin{equation}
	\label{eq:loss:app:2:a}
	\begin{aligned}
		 & \lim_{\tau \to 0^+} \frac1{2\tau} \big\langle h(x_{t+\tau},v_{t+\tau})- h(x_{t-\tau},v_{t-\tau}) \big| (x_t,v_t) =(x,v)\big\rangle           \\
		 & = \lim_{\tau \to 0^+} \frac1{2\tau} \big\langle h(x_{\tau},v_{\tau})- h(x_{-\tau},v_{-\tau}) \big| (x_0,v_0) =(x,v)\big\rangle               \\
		 & = \lim_{\tau \to 0^+} \frac1{2\tau} \big\langle h(x_{\tau},v_{\tau}) - h(x,v) \big| (x_0,v_0) =(x,v)\big\rangle                              \\
		 & \qquad -  \lim_{\tau \to 0^+} \frac1{2\tau} \big\langle h(x^\rev_{\tau},-v^\rev_{\tau}) - h(x,v)\big| (x^\rev_0,v^\rev_0) =(x,-v)\big\rangle \\
		 & = \frac12\big(v\cdot \nabla_x h(x,v) + (f(x,v) - \gamma v ) \cdot \nabla _v h(x,v) + \gamma v_*^2 \Delta_v h(x,v)\big)                       \\
		 & \quad - \frac12 \big( -v\cdot \nabla_x h(x,v) + (f(x,v) - \gamma v -2g(x,v) ) \cdot \nabla _v h(x,v) + \gamma v_*^2 \Delta_v h(x,v) \big)
	\end{aligned}
\end{equation}
Canceling the terms in the last expression gives~\eqref{eq:loss:app:2}.
If we now formally write the limit on the left-hand side of~\eqref{eq:loss:app:2} as a time derivative, we can state this equation as
\begin{equation}
	\label{eq:loss:app:4}
	\begin{aligned}
		\Big\langle \frac{d}{dt}h(x_{t},v_{t}) \Big| (x_t,v_t)=(x,v)\Big\rangle = v \cdot \nabla _x h(x,v) + g(x,v) \cdot \nabla _v  h(x,v).
	\end{aligned}
\end{equation}
Furthermore, if we formally use the chain rule to compute the time derivative in~\eqref{eq:loss:app:4}, we may write
\begin{equation}
	\label{eq:loss:app:3}
	\begin{aligned}
		\Big\langle \frac{d}{dt}h(x_{t},v_{t}) \Big| (x_t,v_t)=(x,v)\Big\rangle & = \big\langle \dot x_t \cdot \nabla_x h(x_{t},v_{t}) + \dot v_t \cdot \nabla_v h(x_{t},v_{t}) \big| (x_t,v_t)=(x,v)\big\rangle \\
		                                                                        & = \big\langle v_t \cdot \nabla_x h(x_{t},v_{t}) + \dot v_t \cdot \nabla_v h(x_{t},v_{t}) \big| (x_t,v_t)=(x,v)\big\rangle      \\
		                                                                        & = v \cdot \nabla_x h(x,v) + \langle \dot v_t | (x_t,v_t)=(x,v)\rangle \cdot \nabla_vh(x,v)
	\end{aligned}
\end{equation}
Comparing~\eqref{eq:loss:app:4} and \eqref{eq:loss:app:3} gives $g(x,v)=\langle \dot v_t | (x_t,v_t)=(x,v)\rangle$.

\section{Macroscopic derivation of the total EPR}
\noindent
To obtain an expression for the macroscopic total EPR, we now consider a path integral derivation~\citep{onsager_fluctuations_1953,machlup_fluctuations_1953} of the change of measure between the two processes~\cref{eqn:vicsek} and~\cref{eqn:reverse_2}.
To this end, we define the measures for a path $\phi_T = \{x_t, v_t\}_{t\in[0, T]}$,
\begin{equation}
	\label{eqn:path_measures}
	\begin{aligned}
		\mathcal{P}_T(\phi_{T})        & = \frac{1}{Z}\exp\left(-\int_{0}^{T}L_{v}(x_{t}, v_{t}, \dot{v}_{t})dt\right)\delta\left(\{\dot{x}_t - v_t\}_{{t \in [0, T]}}\right),        \\
		\mathcal{P}_T^{\rev}(\phi_{T}) & = \frac{1}{Z}\exp\left(-\int_{0}^{T}L_{v}^{\rev}(x_{t}, v_{t}, \dot{v}_{t})dt\right)\delta\left(\{\dot{x}_t - v_t\}_{{t \in [0, T]}}\right),
	\end{aligned}
\end{equation}
along with the Lagrangians
\begin{equation}
	\label{eqn:lagrangians}
	\begin{aligned}
		L_{v}(x_{t}, v_{t}, \dot{v}_{t})        & = \frac{1}{4\gamma v_{*}^{2}}|\dot{v}_{t} - f(x_t, v_t) + \gamma v_t|^{2},                  \\
		L_{v}^{\rev}(x_{t}, v_{t}, \dot{v}_{t}) & = \frac{1}{4\gamma v_{*}^{2}}|\dot{v}_{t} - f(x_t, v_t) + \gamma v_t - 2 g(x_t, -v_t)|^{2}.
	\end{aligned}
\end{equation}
In~\eqref{eqn:path_measures}, the Dirac delta function enforces the constraint $\dot{x}_t = v_t$.
Expanding $L_{v}^{\rev}$, we find that
\begin{equation*}
	\begin{aligned}
		L_v^\rev(x_t, v_t, \dot{v}_t) & = \frac{1}{4\gamma v_0^2}\left(|\dot{v}_t - f(x_t, v_t) + \gamma v_t|^2 - 4 g(x_t, -v_t) \cdot \left(\dot{v}_t - f(x_t, v_t) + \gamma v_t\right) + 4|g(x_t, -v_t)|^2\right), \\
		                              & = L_v(x_t, v_t, \dot{v}_t) - \frac{1}{\gamma v_0^2}g(x_t, -v_t) \cdot \left(\dot{v}_t - f(x_t, v_t) + \gamma v_t\right) + \frac{1}{\gamma v_0^2}|g(x_t, -v_t)|^2.
	\end{aligned}
\end{equation*}
Now, observe that under the forward-time path measure
\begin{equation*}
	\begin{aligned}
		\dot{v}_{t} - f(x_{t}, g_{t}) + \gamma v_t & = \eta_t.
	\end{aligned}
\end{equation*}
It then follows that, with $\langle\cdot\rangle$ denoting an average over the forward-time path measure,
\begin{equation}
	\label{eqn:sdottot}
	\begin{aligned}
		\Sdottot & = \lim_{T\rightarrow\infty}\frac{1}{T}\left\langle\log\left(\frac{\calP_T(\phi_{T})}{\calP_T^{\rev}(\phi_{T})}\right)\right\rangle,                                                               \\
		         & = \lim_{T\rightarrow\infty}\frac{1}{T}\left\langle\int_{0}^{T}\left(L_{v}^{\rev}(x_t, v_t, \dot{v}_t) - L_{v}(x_t, v_t, \dot{v}_t)\right)dt\right\rangle,                                         \\
		         & = \lim_{T\rightarrow\infty}\frac{1}{T}\left\langle\int_{0}^{T}\left(\frac{-1}{\gamma v_0^2}g(x_t, -v_t)\cdot \eta_{t}dt + \frac{1}{\gamma v_{*}^{2}}|g(x_{t}, -v_{t})|^{2}dt\right)\right\rangle, \\
		         & = \lim_{T\rightarrow\infty}\frac{1}{T}\left\langle\int_{0}^{T}\frac{1}{\gamma v_{*}^{2}}|g(x_{t}, -v_{t})|^{2}dt\right\rangle,                                                                    \\
		         & = \frac{1}{\gamma v_{*}^{2}}\int|g(x, -v)|^{2}\rho(x, v)dxdv.
	\end{aligned}
\end{equation}
In the last line, we have used ergodicity of the process to convert the time average to an average over the stationary measure.
We note that by the identity~\cref{eqn:g_gr}, the final line reduces to $\frac{1}{\gamma v_*^2}\langle |g^\rev(x, v)|^{2}\rangle$.

\subsection{Microscopic path measure derivation of the total EPR (continuous)}
\noindent
We now define a stochastic counterpart to the total EPR.
Let $\phi_{t} = (x_\tau, v_\tau)_{\tau \in [0, t]}$ denote a fixed trajectory of the forward system.
Then we consider,
\begin{equation}
	\label{eqn:stot_stoch}
	\begin{aligned}
		s_{\tot}^{\stoch}(t) = \log\left(\frac{\calP_t(\phi_{t})}{\calP_t^{\rev}(\phi_{t})}\right).
	\end{aligned}
\end{equation}
The quantity~\cref{eqn:stot_stoch} measures the probability of observing a \textit{fixed} trajectory $\phi_{t}$ of the \textit{forward system} under the forward and reverse path measures.
As-written, it is an extensive quantity that scales with the horizon $t$.
It is not yet an entropy production rate, but a stochastic measure of entropy that considers how much less likely $\phi_{t}$ is under the reverse dynamics than the forward.
To convert~\cref{eqn:stot_stoch} into a phase-space entropy production rate, we will consider its rate-of-change after averaging over all trajectories $\phi_{t}$ of the forward-time dynamics that end at a given point $(x, v)$ in phase space.
Note that, because we only consider a fixed horizon $t$, this procedure does not cover all possible paths and therefore depends on the final point.
To wit, we consider
\begin{equation}
	\label{eqn:sdottot_stoch}
	\begin{aligned}
		\dot{s}_{\tot}(x, v) = \left\langle\frac{d}{dt}\log\left(\frac{\calP_t(\phi_{t})}{\calP_t^{\rev}(\phi_{t})}\right) \bigg| (x_{t}, v_{t}) = (x, v)\right\rangle.
	\end{aligned}
\end{equation}
From~\cref{eqn:sdottot}, we see that
\begin{equation}
	\label{eqn:sdottot_stoch_algebra}
	\begin{aligned}
		\dot{s}_{\tot}(x, v) & = \left\langle\frac{d}{dt}\int_{0}^{t}\left(L_{v}^{\rev}(x_t, v_t, \dot{v}_t) - L_{v}(x_t, v_t, \dot{v}_t)\right)dt \bigg| (x_{t}, v_{t}) = (x, v)\right\rangle, \\
		                     & = \left\langle L_{v}^{\rev}(x_t, v_t, \dot{v}_t) - L_{v}(x_t, v_t, \dot{v}_t)\bigg| (x_{t}, v_{t}) = (x, v)\right\rangle,                                        \\
		                     & = \left\langle\frac{-1}{\gamma v_*^2}g(x_t, -v_t)\cdot \eta_{t} + \frac{1}{\gamma v_{*}^{2}}|g(x_{t}, -v_{t})|^{2} \bigg| (x_{t}, v_{t}) = (x, v)\right\rangle,  \\
		                     & = \frac{-1}{\gamma v_*^2}g(x, -v)\cdot  \langle\eta_{t}\rangle + \frac{1}{\gamma v_{*}^{2}}|g(x, -v)|^{2},                                                       \\
		                     & = \frac{1}{\gamma v_{*}^{2}}|g(x, -v)|^{2}.
	\end{aligned}
\end{equation}
It then clearly follows that $\langle \sdottot(x, v) \rangle = \Sdottot$.

\subsection{Microscopic path measure derivation of the total EPR (discrete)}
\noindent
In this section, we derive the same result as~\cref{eqn:sdottot_stoch_algebra} from an intuitive discrete-time perspective.
To this end, we consider extending a trajectory $\phi_t$ that ends at $(x_t, v_t) = (x, v)$ infinitesimally, $\phi_{t + \dt} = (\phi_t, (x_{t+\dt}, v_{t+\dt}))$ and explicitly compute its probability under the forward and reverse-time dynamics.
By definition, we have that
\begin{equation}
	\label{eqn:dt_paths}
	\begin{aligned}
		\calP_{t+\dt}(\phi_{t+\dt})        & = \calP_{t}(\phi_{t}) \delta(x_{t+\dt} -x_t - \dt v_{t})\exp\left(-\frac{|v_{t+\dt} -v_t - \dt (f(x_{t}, v_{t}) - \gamma v_{t})|^{2}}{4\gamma v_{*}^{2}\dt}\right), \\
		\calP_{t+\dt}^{\rev}(\phi_{t+\dt}) & = \calP_{t}^{\rev}(\phi_{t})\delta(x_{t+\dt} -x_t- \dt v_{t})                                                                                                       \\
		                                   & \times \exp\left(-\frac{|v_{t+\dt} -v_t - \dt (f(x_{t}, v_{t}) - \gamma v_{t} + 2g(x_{t}, -v_{t}))|^{2}}{4\gamma v_{*}^{2}\dt}\right),
	\end{aligned}
\end{equation}
which may be obtained by taking an Euler step of size $\dt$ of the dynamics~\cref{eqn:vicsek} and~\cref{eqn:reverse_2}.
Equation~\cref{eqn:dt_paths} immediately highlights that to observe $\phi_{t+\dt}$ under the reverse-time dynamics, a very different realization of the noise is required.
Given the functional forms in~\cref{eqn:dt_paths}, we may then explicitly compute
\begin{equation}
	\label{eqn:sdottot_stoch_algebra_discrete}
	\begin{aligned}
		 & \log\left(\frac{\calP_{t+\dt}(\phi_{t+\dt})}{\calP_{t+\dt}^{\rev}(\phi_{t+\dt})}\right) - \log\left(\frac{\calP_{t}(\phi_{t})}{\calP_{t}^{\rev}(\phi_{t})}\right)                                         \\
		 & = \frac{-1}{4\gamma v_{*}^{2}\dt}-|v_{t+\dt} - v_t - \dt (f(x_{t}, v_{t}) - \gamma v_{t})|^{2}                                                                                                            \\
		 & \qquad + \frac{1}{4\gamma v_{*}^{2}\dt}|v_{t + \dt} - v_t - \dt (f(x_{t}, v_{t}) - \gamma v_{t} + 2g(x_{t}, -v_{t}))|^{2},                                                                                \\
		 & = -\frac{1}{4\gamma v_{*}^{2}\dt}\left(|v_{t + \dt} - v_t - \dt (f(x_{t}, v_{t}) - \gamma v_{t})|^{2}\right) + \frac{1}{4\gamma v_{*}^{2}\dt}|v_{t+\dt} - v_t - \dt (f(x_{t}, v_{t}) - \gamma v_{t})|^{2} \\
		 & \qquad + \frac{1}{4\gamma v_{*}^{2}\dt}\left(- 4\dt g(x_{t}, v_{t})\cdot \left(v_{t + \dt} - v_t - dt \cdot \left(f(x_{t}, v_{t}) - \gamma v_{t}\right)\right) + 4\dt^{2}|g(x_{t}, -v_{t}))|^{2}\right),  \\
		 & = \frac{1}{\gamma v_{*}^{2}}\left(\dt|g(x_{t}, -v_{t})|^{2} - g(x_{t}, -v_{t})\cdot\left(v_{t + \dt} - v_t - \dt\left(f(x_{t}, v_{t}) - \gamma v_{t}\right)\right)\right)
	\end{aligned}
\end{equation}
Taking an expectation over the forward-time path $\phi_{t}$ conditioned on the final point yields
\begin{equation}
	\label{eqn:martingale}
	\begin{aligned}
		\left\langle g(x_{t}, -v_{t})\cdot\left(v_{t+\dt} - v_t - \dt\cdot\left(f(x_{t}, v_{t}) - \gamma v_{t}\right)\right) | (x_{t}, v_{t}) = (x, v)\right\rangle & = \sqrt{2 dt \gamma v_{*}^{2}}g(x, -v)\cdot \langle z \rangle \\
		                                                                                                                                                            & = 0,
	\end{aligned}
\end{equation}
where $z \sim \normal(0, I)$ is pure Gaussian noise.
Hence, we find that
\begin{equation}
	\label{eqn:sdottot_stoch_algebra_discrete_2}
	\begin{aligned}
		\left\langle\log\left(\frac{\calP_{t + \dt}(\phi_{t + \dt})}{\calP_{t + \dt}^{\rev}(\phi_{t + \dt})}\right) - \log\left(\frac{\calP_{t}(\phi_{t})}{\calP_{t}^{\rev}(\phi_{t})}\right)\bigg| (x_{t}, v_{t}) = (x, v)\right\rangle & = \frac{\dt}{\gamma v_{*}^{2}}|g(x, -v)|^{2},
	\end{aligned}
\end{equation}
which implies that
\begin{equation}
	\label{eqn:sdottot_stoch_algebra_discrete_3}
	\begin{aligned}
		 & \left\langle\frac{d}{dt}\log\left(\frac{\calP_t(\phi_{t})}{\calP_t^{\rev}(\phi_{t})}\right) \bigg| (x_{t}, v_{t}) = (x, v)\right\rangle                                                                                                                                                       \\
		 & \qquad = \lim_{\dt\rightarrow 0}  \left\langle\frac{1}{\dt}\left[\log\left(\frac{\calP_{t + \dt}(\phi_{t + \dt})}{\calP_{t + \dt}^{\rev}(\phi_{t + \dt})}\right) - \log\left(\frac{\calP_{t}(\phi_{t})}{\calP_{t}^{\rev}(\phi_{t})}\right)\right]\bigg| (x_{t}, v_{t}) = (x, v)\right\rangle, \\
		 & \qquad = \frac{1}{\gamma v_{*}^{2}}|g(x, -v)|^{2},
	\end{aligned}
\end{equation}
as shown in~\cref{eqn:sdottot_stoch_algebra}.

\section{Further details on the loss function}
\noindent
Here, we show that $g(x, v)$ is the unique minimizer of~\cref{eqn:strato_loss} for any $T > 0$.
To this end, the loss reads
\begin{equation}
	\label{eq:loss:app}
	\begin{aligned}
		\calL(\hat{g}) & = \frac{1}{T}\E\left[\int_{0}^{T}|\hat{g}(x_{t}, v_{t})|^{2}dt - 2\hat{g}(x_{t}, v_{t})\circ dv_{t}\right],                                                                                                \\
		               & = \frac{1}{T}\E\left[\int_{0}^{T}|\hat{g}(x_{t}, v_{t})|^{2}dt - 2\hat{g}(x_{t}, v_{t})\cdot \left(f(x_t, v_t) - \gamma v_t\right)dt - 2 \sqrt{2\gamma v_{*}^{2}}\hat{g}(x_{t}, v_{t})\circ dW_{t}\right],
	\end{aligned}
\end{equation}
where we have simply used the equation for $v_t$ in~\cref{eqn:vicsek} to expand the expression for $dv_t$.
Now, converting the Stratonovich integral to an Ito integral and using the fact that Ito integrals are mean zero,
\begin{equation}
	\begin{aligned}
		\calL(\hat{g}) & = \frac{1}{T}\E\left[\int_{0}^{T}|\hat{g}(x_{t}, v_{t})|^{2}dt - 2\hat{g}(x_{t}, v_{t})\cdot \left(f(x_t, v_t) - \gamma v_t\right)dt\right]                                                              \\
		               & \qquad + \frac{1}{T}\E\left[\int_{0}^{T}- 2 \gamma v_{*}^{2}\nabla_{v}\cdot \hat{g}(x_{t}, v_{t})dt - 2 \sqrt{2\gamma v_{*}^{2}}\hat{g}(x_{t}, v_{t})\cdot dW_{t}\right],                                \\
		               & = \frac{1}{T}\E\left[\int_{0}^{T}|\hat{g}(x_{t}, v_{t})|^{2}dt - 2\hat{g}(x_{t}, v_{t})\cdot \left(f(x_t, v_t) - \gamma v_t\right)dt - 2 \gamma v_{*}^{2}\nabla_{v}\cdot \hat{g}(x_{t}, v_{t})dt\right].
	\end{aligned}
\end{equation}
Recognizing that the density of $(x_t, v_t)$ is $\rho$ for all $t$ because we consider dynamics at stationarity, we may drop the time integral.
Integrating the resulting expression by parts and using~\cref{eqn:g_score}, we find
\begin{equation}
	\label{eqn:loss_proof}
	\begin{aligned}
		\calL(\hat{g}) & = \int \left(|\hat{g}(x, v)|^{2} - 2\hat{g}(x, v)\cdot \left(f(x_t, v_t) - \gamma v_t\right) - 2 \gamma v_{*}^{2}\nabla_{v}\cdot \hat{g}(x, v)\right)\rho(x, v)dxdv,                \\
		               & = \int \left(|\hat{g}(x, v)|^{2} - 2\hat{g}(x, v)\cdot \left(f(x_t, v_t) - \gamma v_t\right) + 2 \gamma v_{*}^{2}\hat{g}(x, v) \cdot \nabla_{v}\log\rho(x, v)\right)\rho(x, v)dxdv, \\
		               & = \int \left(|\hat{g}(x, v)|^{2} - 2\hat{g}(x, v)\cdot g(x, v)\right)\rho(x, v)dxdv.
	\end{aligned}
\end{equation}
The final line above is a strongly convex square-loss regression problem with unique minimizer $\hat{g}(x, v) = g(x, v)$ $\rho$-almost everywhere, as may readily be verified by taking the first and second variation with respect to $\hat{g}$.

\section{Expression for $\sdotsys(x, v)$}
\noindent
Applying~\eqref{eq:loss:app:4} with $h(x,v) = \rho(x,v)$, we obtain
\begin{equation}
	\label{eq:sdotsys:app}
	\begin{aligned}
		\sdotsys(x, v) & = -\left\langle \frac{d}{dt}\log\rho(x_t, v_t) \Big| (x_t, v_t) = (x, v) \right\rangle \\
		               & = -v\cdot \nabla_x \log \rho(x, v)- g(x,v) \cdot \nabla_v \log \rho (x, v).
	\end{aligned}
\end{equation}
Now observe that by the stationary FPE in transport form,
\begin{equation}
	\begin{aligned}
		\nabla_x \cdot \left(v \rho(x, v)\right) + \nabla_v \cdot \left(g(x, v) \rho(x, v)\right) = 0,
	\end{aligned}
\end{equation}
so that, after expansion and division by $\rho(x, v)$,
\begin{equation}
	\begin{aligned}
		v \cdot \nabla_x\log\rho(x, v) + \nabla_v \cdot g(x, v) + g(x, v) \cdot \nabla_v\log\rho(x, v) = 0.
	\end{aligned}
\end{equation}
Solving this equation for $\nabla_v \cdot g(x, v)$ and inserting the result in~\eqref{eq:sdotsys:app} we find that
\begin{equation}
	\sdotsys(x, v) = -\left\langle \frac{d}{dt}\log\rho(x_t, v_t) \Big| (x_t, v_t) = (x, v) \right\rangle = \nabla_v \cdot g(x, v).
\end{equation}

\section{Vicsek system details}
\noindent
In this section, we report some additional details on the Vicsek system studied in the main text, and report a few additional results for the low-dimensional system.

\subsection{Kernel}
\noindent
We set the kernel $K$ to be a smooth variant of the Heaviside function,
\begin{equation}
	\label{eqn:sigmoid}
	\begin{aligned}
		K(x^{i} - x^{j}) & = \sigma\left(-\beta \left(\norm{x^{i}-x^{j}}_{2}^{2} - 4r^{2}\right)\right), \\
		\sigma(x)        & = \frac{1}{1 + e^{-x}}.
	\end{aligned}
\end{equation}
In~\eqref{eqn:sigmoid}, $r$ denotes the interaction radius of the particles and $\beta$ is a smoothing parameter.
As $\beta\rightarrow \infty$, the Heaviside function $\Theta(2r - \norm{x^{i}-x^{j}}_{2})$ is recovered.
We find that the qualitative physical dynamics are unaffected by the smoothing used in~\cref{eqn:sigmoid}, but that the learning problem simplifies, as it ensures that the neural network does not need to learn a discontinuous function.
We choose $r$ so that the packing fraction $\phi = 0.5$ on the box $[-1, 1]^2$.

\subsection{Low-dimensional system}
\noindent
Here, we derive the equivalent low-dimensional system in displacement coordinates studied in the main text.
The system~\eqref{eqn:vicsek} with two particles in dimension $d=1$ reads,
\begin{equation}
	\label{eqn:vicsek_2p1d}
	\begin{aligned}
		dx^1_t & = v^1_t dt,                                                                           \\
		dx^2_t & = v^2_t dt,                                                                           \\
		dv^1_t & = -\gamma v^1_tdt + (v^2_t - v^1_t)K(x^1_t - x^2_t)dt + \sqrt{2\gamma v_*^2}dW^1_{t}, \\
		dv^2_t & = -\gamma v^2_tdt + (v^1_t - v^2_t)K(x^1_t - x^2_t)dt + \sqrt{2\gamma v_*^2}dW^2_{t},
	\end{aligned}
\end{equation}
with $x^i_t, v^i_t \in \R$.
Defining the displacement coordinates $x_t = x^1_t - x^2_t$ and $v_t = v^1_t - v^2_t$, we obtain a two-dimensional system whose phase space can be visualized,
\begin{equation}
	\label{eqn:vicsek_lowd}
	\begin{aligned}
		dx_t & = v_t dt,                                                         \\
		dv_t & = -\left(\gamma + 2K(x)\right)v_tdt + 2\sqrt{\gamma v_0^2}dW_{t}.
	\end{aligned}
\end{equation}
In practice, we simulate~\eqref{eqn:vicsek_2p1d} and learn from the two-particle data, but then convert to~\eqref{eqn:vicsek_lowd} for visualization.

\subsubsection{Network architecture and training}
\noindent
For this low-dimensional system, we learn usingwa simple four-layer fully-connected network with $512$ neurons in each layer and a GeLU activation.
The network enforces the $(x, v) \mapsto (-x, -v)$ symmetry of the system by taking $g(x, v) = \tfrac{1}{2}(N(x, v) - N(-x, -v))$ where $N$ is the fully-connected network.

\subsubsection{Additional results}
\begin{figure}[!tbh]
	\centering
	\begin{overpic}[width=0.5\textwidth]{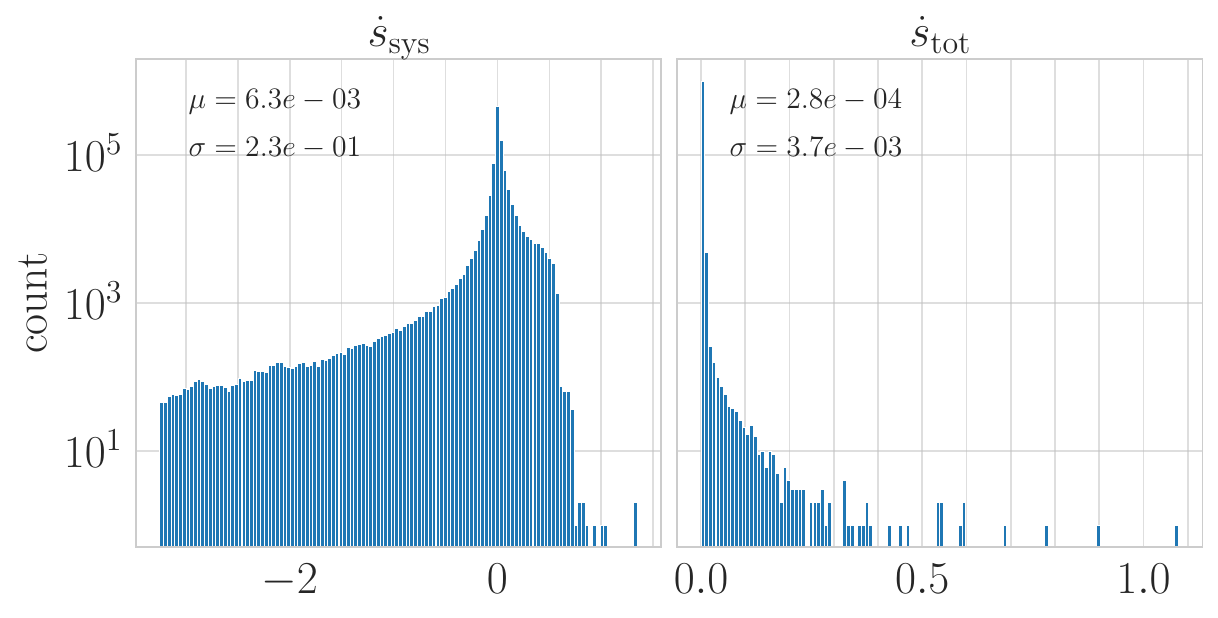}
		\put(7.5, 48){\textbf{A}}
		\put(54,  48){\textbf{B}}
	\end{overpic}
	\caption{
		\textbf{Two Vicsek particles in one dimension: statistics of the stationary distribution}.
		(A) Density of the microscopic system EPR. A heavy negative tail indicates asymmetry between events of large entropy production and events of large entropy consumption.
		(B) Density of the microscopic total EPR. The distribution has as heavy tail generated by collision events between the particles.
	}
	\label{fig:lowd_vicsek_statistics}
\end{figure}

\begin{figure*}[!t]
	\includegraphics[width=\textwidth]{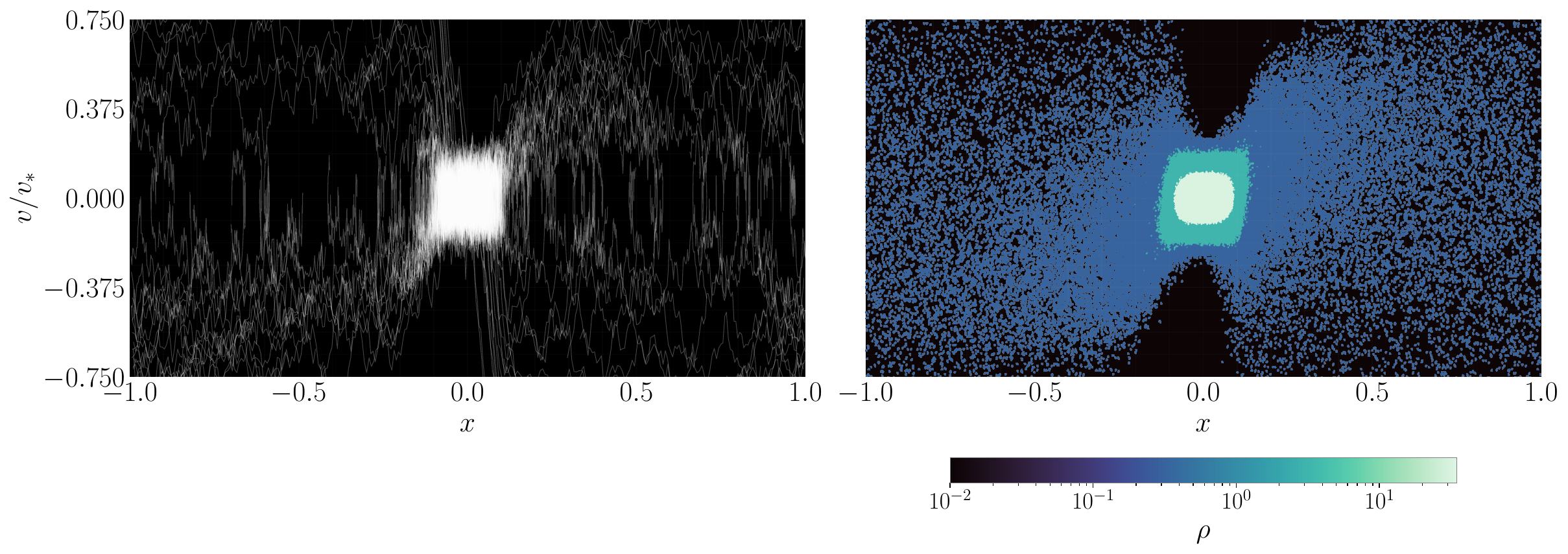}
	\caption{
		(Left) Example trajectories from the SDE~\cref{eqn:vicsek_lowd}.
		(Right) Stationary density for the SDE~\cref{eqn:vicsek_lowd}, obtained by computing a histogram from samples.
	}%
	\label{fig:vicsek_lowd_sde_density}
\end{figure*}

\noindent
Statistics of the learned quantities on the NESS distribution are shown in~\cref{fig:lowd_vicsek_statistics}.
These statistics summarize what can be seen visually over the phase space in~\cref{fig:vicsek_lowd_EPR}.
The system EPR $\sdotsys$ is asymmetric, and has a heavy negative tail, highlighting that entropy production and entropy consumption occur via distinct mechanisms.
Nevertheless, as required by the stationarity condition $\partial_t \langle \int \log\rho(x, v) \rho(x, v)dxdv \rangle$, we have that $\langle\sdotsys(x, v)\rangle \approx 0$.
The statistics of the learned $\sdottot(x, v)$ also have a heavy tail, generated via the collision mechanism discussed in the main text.
A collection of stochastic trajectories, along with the associated NESS distribution, is shown in~\cref{fig:vicsek_lowd_sde_density}.
The trajectories concentrate around the aligned state, but generate a slight asymmetry that can be seen in the NESS distribution.

